\documentclass[fleqn,12pt,twoside]{article}
\usepackage[headings]{espcrc1}

\readRCS
$Id: espcrc1.tex,v 1.2 2004/02/24 11:22:11 spepping Exp $
\usepackage{graphicx}
\usepackage[figuresright]{rotating}

\newcommand{\AmS}{{\protect\the\textfont2
  A\kern-.1667em\lower.5ex\hbox{M}\kern-.125emS}}

\title{Toward a Global Dispersive Optical Model for the Driplines}

\author{C. Barbieri\address{Theoretical Nuclear Physics Laboratory, \\
                   RIKEN Nishina Center, 2-1 Hirosawa, Wako, Saitama 351-0198, Japan},
R. J. Charity\address[wuchem]{Department of Chemistry, Washington University, St. Louis, Missouri 63130, USA},
W. H. Dickhoff\address[wuphys]{Department of Physics, Washington University, St. Louis, Missouri 63130, USA}
and
L. G. Sobotka\addressmark[wuchem] \addressmark[wuphys]}
       
\runtitle{DOM for Ca isotopes}
\runauthor{C. Barbieri, et al.}

\begin{document}

\maketitle

\begin{abstract}
A dispersive-optical-model analysis has been performed for both
protons and neutrons on $^{40,42,44,48}$Ca isotopes. The fitted potentials
describe accurately both scattering and bound quantities and extrapolate
well to other stable nuclei.
 Further experimental information will be gathered to constrain
extrapolations toward the driplines.
\end{abstract}

\section{Introduction}

Experiments with radioactive beams are unraveling the properties of
nuclei at the limits of stability~\cite{exotic}.
 Studies of low-energy processes and transfer reactions
require optical potentials~(OP) suitable for dripline isotopes.
This is also important for astrophysics applications, where capture
and decay processes require a consistent determination of both
scattering and bound states. 
 Dispersive optical models (DOM) provide a useful tool that allows
one to describe the mean field consistently over a  broad energy range.

The DOM is grounded in the theory of
many-body Green's functions (or propagators)~\cite{dicvan}
and in the proven equivalence between the many-body self-energy
(of propagator theory) and the Feshbach theory of the microscopic
optical model~\cite{Cap.96}.
In practice the DOM can be thought as a parametrization of the
self-energy. This has several advantages. It allows us to develop
an OP that describes consistently both elastic
scattering and bound-state properties.
The theory directly imposes important analytic \hbox{constraints,
such as the dispersion} relation between the real and imaginary
components (see below).
Also microscopic calculations~\cite{Bar.05} can be employed
to determine the physical ingredients of the potentials. 

The DOM was 
developed by Mahaux and collaborators for a number of
individual magic or near magic nuclei~\cite{Mah.91}.
Recently, the authors of Refs.~\cite{Cha.06,Cha.07} have 
performed a DOM study for the chain of Ca isotopes. This employed
the world data on $^{40-48}$Ca,
including {\em both} elastic scattering (up to 200~MeV) and 
bound-state information from ($e$,$e'p$) experiments.
 This analysis also discussed possible extrapolations toward
the driplines and suggested new experiments to constrain the
predictions for exotic nuclei~\cite{Ca48exp}.
A global parametrization of the coupled-channel DOM was also reported
in Ref.~\cite{Hao.08} for stable nuclei.

 The present talk discusses briefly the analysis of Refs.~\cite{Cha.06,Cha.07} and
points to open issues that are being addressed to refine the predictions
for unstable isotopes.

\begin{figure}[t]
\includegraphics[width=0.85\textwidth]{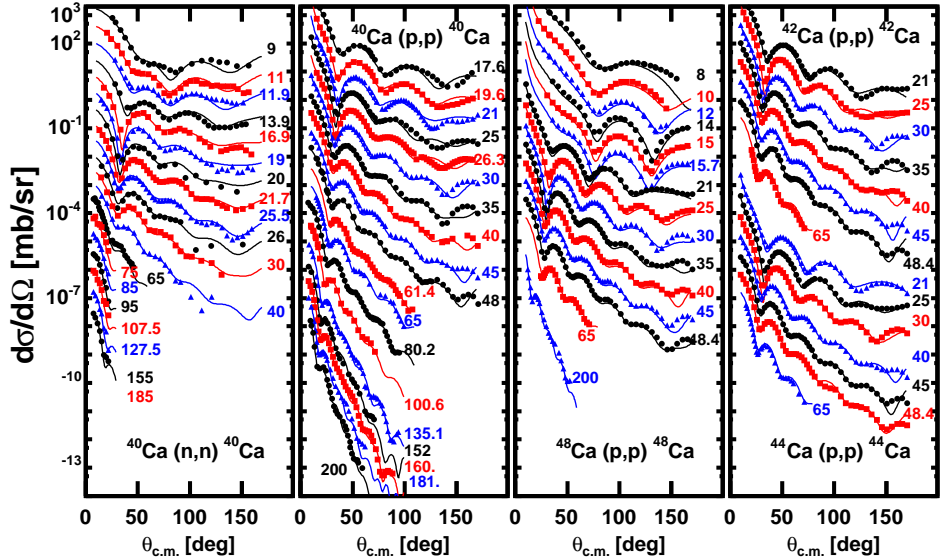}
\vskip -.3cm
\caption{(Color online) Comparison of experimental (data) and fitted (curves)
differential elastic scattering cross sections. Successively larger energies
are scaled down by a factor of 4. For p+$^{42}$Ca, the data and curves
are scaled down by an additional factor of 100.}
\label{fig:ds}
\end{figure}

\section{The DOM for Ca Isotopes}

While the nucleon self-energy is an {\em energy dependent} and
{\em non-local} potential, the vast majority of DOM applications employ a
local approximation of it.  The same approach is followed here, although
this approximation may be removed in future works. 
 Mahaux and Sartor~\cite{Mah.91} start by separating the real part of the
optical potential
 at the Fermi surface (${\cal V}_{HF}$) and write the DOM as
\begin{equation}
 {\cal U}(r,E) = {\cal V}_{HF}(r,E) + \Delta{\cal V}(r,E) + i{\cal W}(r,E)
\; ,
\label{eq:DOM}
\end{equation}
where ${\cal V}_{HF}(r,E)$ is the local equivalent of the self-energy,
$\textrm{Re} \Sigma(r,r';E_F)$, at the Fermi energy ($E_F$)%
\footnote{Strictly speaking, ${\cal V}_{HF}$ is not an Hartree-Fock potential
but it does describe the effects of the nuclear mean-field. We maintain this
notation for consistency with the rest of the literature.}.
Normally, this would be a static potential but it acquires an energy-dependence
to account for the non-locality of $\Sigma(r,r';E_F)$.
The real and imaginary parts of the dynamic components of the DOM are 
linked by the subtracted {\em dispersion relation},
\begin{equation}
 \Delta{\cal V}(r,E) = \frac 1 \pi P \int {\cal W}(r,\omega) 
 \left( \frac 1{\omega - E} - \frac 1{\omega - E_F} \right)
 d \omega
\; .
\end{equation}
This relation is a direct consequence of the causality principle,
which is therefore embedded in the model. 
Thus, the DOM is fully determined by parametrizing only the
${\cal V}_{HF}(r,E)$ and ${\cal W}(r,E)$ components. The explicit 
form employed in this work is reported in detail in Ref.~\cite{Cha.07}.

\begin{figure}[t]
\begin{minipage}[t]{93mm}
\includegraphics[width=1\textwidth]{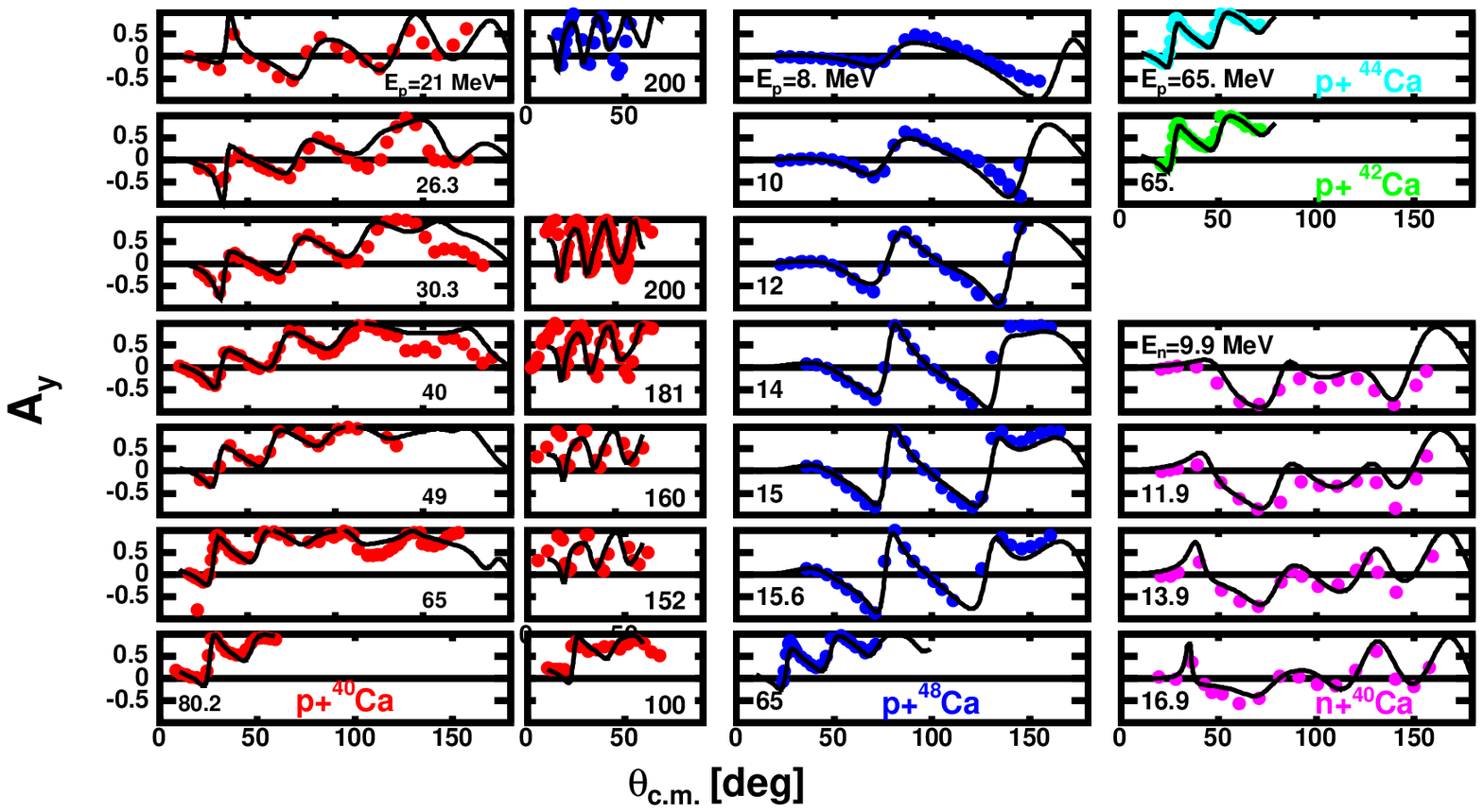}
\end{minipage}
\hspace{\fill}
\begin{minipage}[t]{52mm}
\includegraphics[width=1\textwidth]{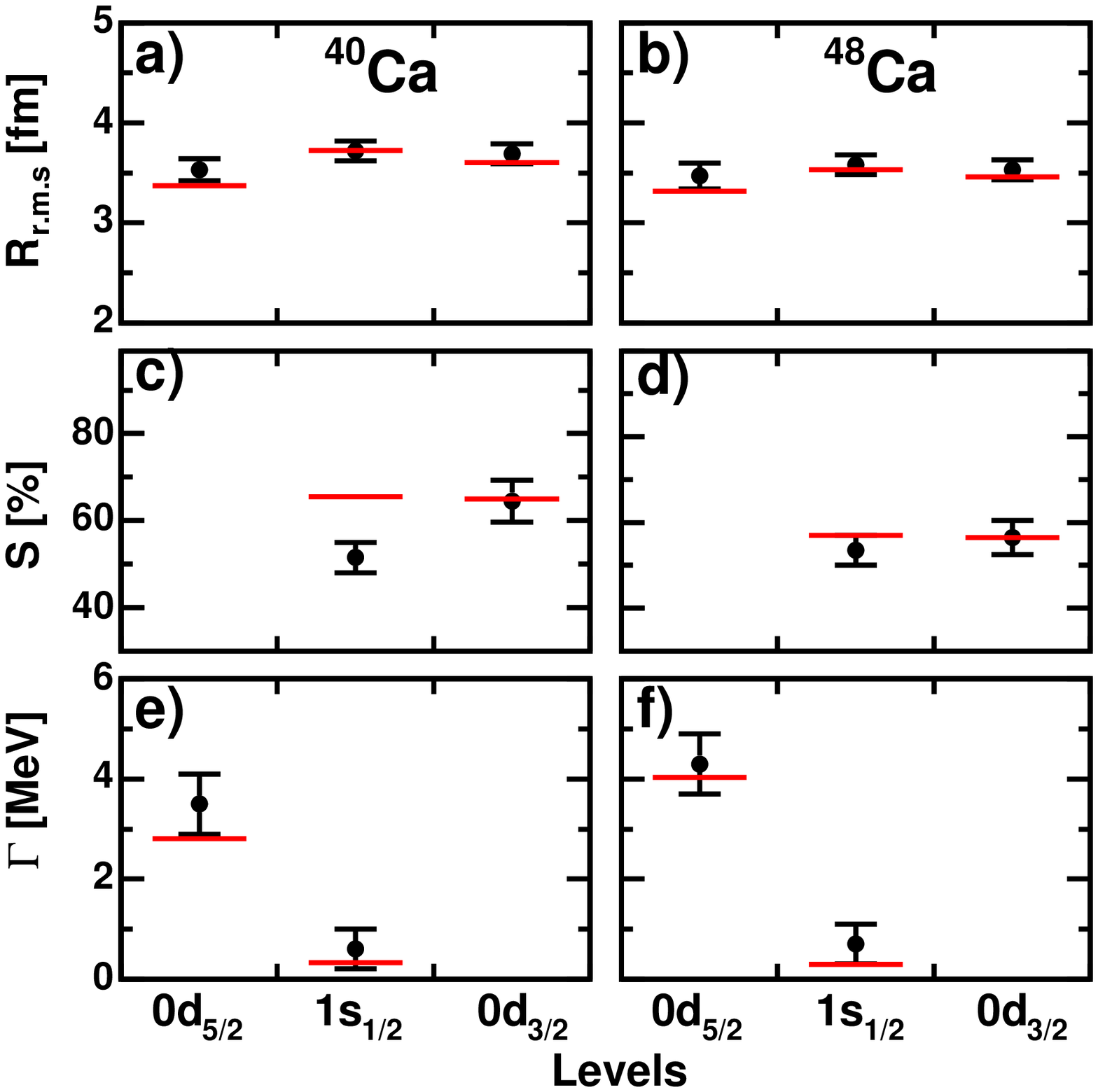}
\end{minipage}
\label{fig:Ay}
\vskip -0.3cm
\caption{(Color online) {\em Left.} Comparison of experimental and
fitted analyzing powers.
{\em Right.}  Fitted (horizontal lines) and experimental
(data points) level properties for the 0$d_{5/2}$, 1$s_{1/2}$,
and 0$d_{3/2}$ proton hole states for $^{40}$Ca~(left panels) and
$^{48}$Ca~(right panels). The
fitted quantities include the root-mean-squared radius $R_{rms}$,
the spectroscopic factors $S$ and the widths
$\Gamma$ of these states.}
\end{figure}

The potential~(\ref{eq:DOM}) was fitted to published data sets
at both positive and negative energies. These included elastic scattering
data for $p+^{40,42,44,48}$Ca
and $n+^{40}$Ca, with energies up to 200~MeV, total reaction cross sections,
experimental single-particle levels and spectroscopic factors and radii
of proton bound states measured in ($e$,$e'p$) reactions.
The quality of the fit to the differential cross sections and analysing powers
is shown in Figs.~\ref{fig:ds} and~\ref{fig:Ay}~(left). 
For proton bound states, only one data point could not be reproduced, i.e.,
the $s_{1/2}$ orbit in $^{40}$Ca as shown in Fig.~\ref{fig:ds}~(right).
 An independent re-analysis of ($e$,$e'p$) data has later corrected this
value and resolved this discrepancy~\cite{Lap.s12}.

For such a large body of data (81 data sets comprising 3569 data points),
the excellent agreement of the fit with just 25 free parameters provides
confidence in the predictive power of the DOM calculations.

\section{Extrapolation to Other Isotopes}

\begin{figure}[htb]
\vskip -.3cm
\includegraphics[width=0.9\textwidth]{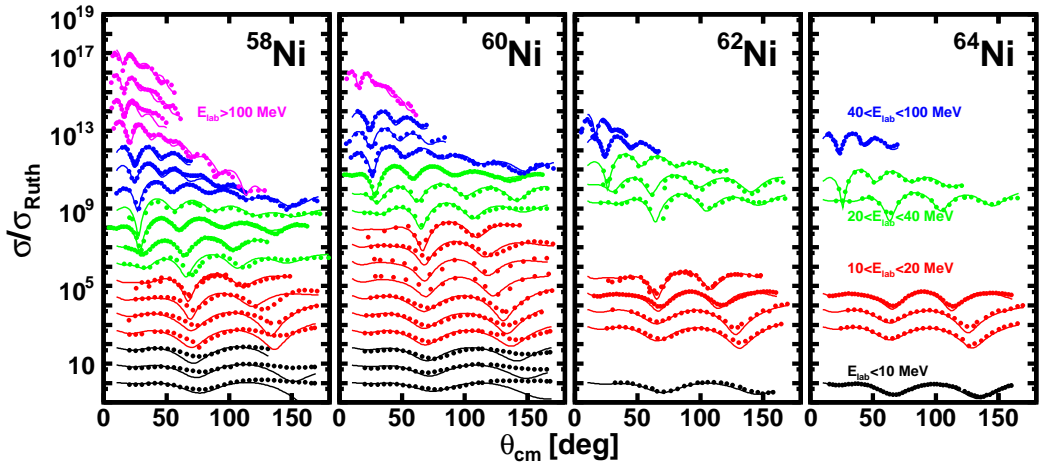}
\caption{(Color online) {\em Preliminary} comparison between the proton
scattering data on Ni isotopes and the prediction from the DOM
of Ref.~\cite{Cha.07} (fitted solely to Ca isotopes). The plot shows
the ratio with respect to the Rutherford differential cross sections.}
\label{fig:Ni}
\end{figure}

Figure~\ref{fig:Ni} shows a {\em preliminary} comparison between the
elastic proton scattering on stable Ni sotopes and the prediction
of the above DOM, which was fitted to Ca data only.  The overall agreement 
is quite satisfying and shows the potentiality for global fits 
with this approach.
However, proper extrapolations to unstable isotopes (and
in particular close to the driplines) require particular
care for the N/Z asymmetry dependence of the surface part
of the imaginary potential. This is usually assumed to behave as
\begin{equation}
 W_s(E) = W_s^0(E) + \varepsilon \frac{N-Z}{A} W_s^1(E) 
\; .
\label{eq:asymm}
\end{equation}
The choice for protons is $\varepsilon_p$=+1, which works 
quite well for $p$+$^{40-48}$Ca scattering.  For neutrons, the standard
assumption in the literature ($\varepsilon_n$=-1)
leads to inconsistencies for neutron-rich isotopes. Similar difficulties
have also been reported in a recent global OP fit~\cite{Wep.09}.

Unfortunately, the n+$^{40}$Ca scattering data solely available to
Ref.~\cite{Cha.07} are not sufficient to constrain $\varepsilon_n$.
Choosing $\varepsilon_n$=0 was found to be consistent with
spectroscopic factors from ($d$,$p$) reactions but does not rule
out other possibilities.
 New data for the n+$^{48}$Ca have recently been taken to verify this
assumption~\cite{Ca48exp}.  It is expected that this will allow new fits
that improve the description of dripline isotopes.

\vskip .2cm

This work is supported by the U.S. Department of Energy, Division of Nuclear 
Physics under grant DE-FG02-87ER-40316 and the U.S. National Science Foundation 
under grant PHY-0652900. C.B. acknowledges a KAKENHI grant (no. 21740213) 
form the Japanese MEXT.


\begin{thebibliography}{9}

\bibitem{exotic}  A.~Gade and T.~Glasmacher, Prog. Part. Nucl. Phys. {\bf 60}, 161 (2008).
\bibitem{dicvan}  W.~H.~Dickhoff and D.~Van~Neck, {\em Many-Body Theory Exposed!} (2nd edition, World Scientific, Singapore, 2008).
\bibitem{Cap.96}  F.~Capuzzi and C.~Mahaux, Ann. Phys. {\bf 245}, 147 (1996).
\bibitem{Bar.05}  C. Barbieri and and B. K. Jennings, Phys. Rev. C {\bf 72}, 014613 (2005).
\bibitem{Mah.91}  C.~Mahaux and R.~Sartor, Adv. Nucl. Phys. {\bf 20}, 1 (1991).
\bibitem{Cha.06}  R.~J.~Charity, L.~G.~Sobotka, W.~H.~Dickhoff, Phys.\ Rev.\ Lett. {\bf 97}, 162503 (2006).
\bibitem{Cha.07}  R.~J.~Charity, J.~M.~Mueller, L.~G.~Sobotka, W.~H.~Dickhoff, Phys.\ Rev.\ C {\bf 76}, 044314 (2007).
\bibitem{Ca48exp} R.~J.~Charity, L.~G.~Sobotka {\em et al.}, in preparation.
\bibitem{Hao.08}  L.~Hao, W.~Sun, E.~Sh.~Soukhovitskii, J. Phys. G {\bf 35} 095103 (2008).
\bibitem{Lap.s12} L.~Lapik\'{a}s, priv. comm. and to be published.
\bibitem{Wep.09}  S.~P.~Weppner, Phys. Rev. C {\bf 80}, 034608 (2009).


\end{thebibliography}
\end{document}